\documentclass[twocolumn,aps]{revtex4-1}

\usepackage[english]{babel}
\usepackage{amsmath}
\usepackage{latexsym}
\usepackage{amssymb}

\usepackage{graphicx}

\usepackage{amsmath} 
\usepackage{color}

\usepackage{mathptmx}      

\newcommand{\mH}{\mathcal{H}}
\newcommand{\mI}{\mathcal{I}}

\newcommand{\mP}{\textcolor{black}{P}}
\newcommand{\mQ}{\textcolor{black}{Q}}

\newcommand{\mZ}{\mathcal{Z}}
\newcommand{\mU}{\mathcal{U}}

\newcommand{\be}{\begin{equation}}
\newcommand{\ee}{\end{equation}}
\newcommand{\bea}{\begin{eqnarray}}
\newcommand{\eea}{\end{eqnarray}}
\newcommand{\bse}{\begin{subequations}}
\newcommand{\ese}{\end{subequations}}
\newcommand{\comment}[1]{}

\newcommand{\ggcol}[1]{{\color{black} #1}}
\newcommand{\mcol}[1]{{\color{black} #1}}

\begin{document}

\title{Statistical mechanics of an integrable system}

\author{Marco Baldovin}
\affiliation{Dipartimento di Fisica, Universit\`a di Roma ``Sapienza'', Piazzale A. Moro 2, I-00185, Roma, Italy}

\author{Angelo Vulpiani}
\affiliation{Dipartimento di Fisica, Universit\`a di Roma ``Sapienza'', Piazzale A. Moro 2, I-00185, Roma, Italy}

\author{Giacomo Gradenigo}
\affiliation{Gran Sasso Science Institute, Viale F. Crispi 7, 67100 L'Aquila, Italy}
\email{giacomo.gradenigo@gssi.it}


\begin{abstract}

  We provide here an explicit example of Khinchin's idea that the
  validity of equilibrium statistical mechanics in high dimensional
  systems does not depend on the details of the dynamics, as it is
  basically a matter of choosing the ``proper'' observables.  This
  point of view is supported by extensive numerical simulation of the
  one-dimensional Toda chain, an integrable non-linear Hamiltonian
  system where all Lyapunov exponents are zero by definition.  We
  study the relaxation to equilibrium starting from very atypical
  initial conditions and focusing on energy equipartion among Fourier
  modes, as done in the original Fermi-Pasta-Ulam-Tsingou numerical
  experiment. We consider other indicators of thermalization as well,
  e.g. Boltzmann-like probability distributions of energy and the
  behaviour of the specific heat as a function of temperature, which
  is compared to analytical predictions. We find evidence that in the
  general case, i.e., not in the perturbative regime where Toda and
  Fourier modes are close to each other, there is a fast reaching of
  thermal equilibrium in terms of a single temperature.  We also find
  that equilibrium fluctuations, in particular the behaviour of
  specific heat as function of temperature, are in agreement with
  analytic predictions drawn from the ordinary Gibbs ensemble. The
  result has no conflict with the established validity of the
  Generalized Gibbs Ensemble for the Toda model, which is on the
  contrary characterized by an extensive number of different
  temperatures.  Our results suggest thus that even an integrable
  Hamiltonian system reaches thermalization on the constant energy
  hypersurface, provided that the considered observables do not
  strongly depend on one or few of the conserved quantities. This
  suggests that dynamical chaos is irrelevant for thermalization in
  the large-$N$ limit, where any macroscopic observable reads of as a
  collective variable with respect to the coordinate which diagonalize
  the Hamiltonian. The possibility for our results to be relevant for
  the problem of thermalization in generic quantum systems, i.e.,
  non-integrable ones, is commented at the end.

  
\end{abstract}

\maketitle

\section{Introduction}

It is common wisdom that statistical mechanics works rather well at a
practical level, meaning that its predictions about the relevant
observables of a macroscopic system, based on equilibrium ensemble
averages, correctly describe the actual behavior measured in
experiments or numerical simulations~\cite{MA85,JS06}.  \ggcol{For
  instance, statistical mechanics allows us to understand the features
  of macroscopic objects, including intriguing phenomena such as
  criticality and scale invariance. It is then quite natural to wonder
  about the origin of such a great success.\\
  
  The problem of statistical mechanics foundations in weakly
  non-linear Hamiltonian systems has been widely investigated. A
  paradigmatic and historically crucial example is the
  Fermi-Pasta-Ulam-Tsingou (FPUT)
  problem~\cite{FPU55,LPRSV85,G05,FIK05,BCP13}; many works originated
  by the study of this model showed that there are two main competing
  ingredients which allow for the emergence of a thermal phase in
  non-linear high-dimensional systems: \textit{(i)} dynamical chaos,
  favouring thermalization, and \textit{(ii)} the formation of stable
  or metastable recurrent excitations (breathers, solitons, etc.),
  which on the contrary slow down or completely arrest the relaxation
  to a thermal state~\cite{F73,F92,CRZ05}.
  
 This said, the true relevance of chaos to legitimate statistical
 mechanics is still an open issue.}  Roughly speaking the different
points of view may be traced back to two opposite schools:
\textit{(i)} the ``chaotic'' one (based on the work of Prigogine and
of his followers~\cite{PG98}), according to which a key role is played
by the dynamics and, consistently, the presence of chaos is regarded
as the basic ingredient for the validity of the statistical mechanics;
\textit{(ii)} the ``traditional'' one, following the original
Boltzmann's ideas~\cite{K49,CFLV08,B95,L93,Z05}, which stresses the
role of the large number of degrees of freedom. Important
contributions to the latter point of view are due to
Khinchin~\cite{K49}, as well as to Mazur and van der
Linden~\cite{ML63}. Their results show that, although for most
physical systems the ergodic hypothesis does not hold
\emph{mathematically}, it is \emph{physically} valid in the following
sense: if we limit our interest to collective variables, in systems
with a large number of degrees of freedom $N$, the measure of the
regions where ergodicity fails vanishes in the limit $N \to \infty$.
In this respect the dynamics plays a marginal role, the key ingredient
being only the assumption $N \gg 1$ and the use of collective
observables, i.e., functions of \emph{all} canonical coordinates (or,
at least, of a finite fraction of them). \ggcol{Khinchin's approach
  has some practical limitations, in particular it is not constructive
  from the point of view of finding criteria to assess
  thermalization. For a given setup, i.e., a given choice of the
  observable and the initial conditions, it gives no hints about
  thermalization timescales.}  Despite these difficulties, Khinchin's
point of view still seems to us the most promising way to gain some
insight into \ggcol{ the reason why macroscopic systems do actually
  thermalize.}

To show this point, in this paper we consider a well known integrable
Hamiltonian system, whose dynamics is, by definition, non chaotic.  We
will show that, even in this case, when thermalization indicators are
measured with respect to canonical variables different from those
which diagonalize the Hamiltonian, thermal equilibrium is reached
within very reasonable times.  Our numerical results strongly support
the following agnostic scenario:~\emph{thermalization} is not an
intrinsic property of the system but rather a property of the chosen
description. \ggcol{It is from this point of view that our results are
  not in conflict with recent results on the Toda lattice~\cite{S20},
  according to which the thermodynamics of this system is described by
  the Generalized Gibbs Ensemble (GGE). GGE is a Gibbs-like ensemble,
  initially introduced for isolated quantum
  systems~\cite{RMO06,RDYO07}, where to each conservation law of the
  Hamiltonian dynamics is associated a different temperature. In
  practice in the GGE the partition function reads as $\mZ =
  \text{Tr}[e^{-\sum_{i=1}^N \beta_i Q_i}]$, where the $Q_i$'s
  represent the quantities conserved by the dynamics, i.e.,
  classically, $\left\lbrace Q_i,\mH\right\rbrace=0$, with $\left\lbrace
  \cdot , \cdot \right\rbrace$ being the Poisson parentheses. Although
  this ensemble provided consistent predictions for correlation
  functions in integrable models, initially with a big emphasis on
  quantum systems~\cite{RMO06,RDYO07,RDO08,RSMS09,CEF11,VR16} and more
  recently also on classical
  ones~\cite{S20,D90,dLM16,GMI19,SBDD19,dVBdLM20}, we believe that it
  is just a particular choice to describe the system. In fact, there
  are as many ways to write the partition function of a Hamiltonian
  system as are the possible choices for canonically conjugated
  variables. GGE represents a natural choice to emphasize the
  existence of many time scales and many energy scales in an
  integrable system. But is the knowledge of all these time and energy
  scales really necessary for the calculation of relevant macroscopic
  properties of the system? In particular is the GGE really necessary
  to estimate, just to make an example, the time average of global
  observables which depend, in general, on an extensive number of
  conserved quantities?  These are in our opinion open questions which
  call for further investigations. In particular, the contribution of
  this paper is the presentation of neat numerical data which suggest
  that thermalization occurs at a single temperature for generic
  macroscopic observables, and their comparison with analytical
  calculations in the standard Gibbs ensemble. It is worth mentioning
  that in small systems even generic observables keep a strong
  correlation with conserved quantities, see for instance~\cite{O15}:
  in this respect the large-$N$ limit of the Khinchin approach seems
  to be crucial. Focusing on classical integrable systems is from this
  point of view particularly convenient since much larger sizes,
  compared to the quantum case, can be tested numerically.\\
  
  The idea that the possibility to achieve thermalization depends on
  the choice of the observed degrees of freedom was proposed for
  instance in the domain of disordered systems with ergodicity
  breaking transitions: it is well known that in the
  replica-symmetry-broken phase of spin-glasses some degrees of
  freedom reach fast equilibration with the thermostat whereas others
  never equilibrate~\cite{CKP97}. And this idea was proposed already
  before in~\cite{M69,PC77}. Incidentally, from this point of view
  intriguing similarities between spin glasses and the GGE have been
  recently put in evidence in a series of papers focused on the idea
  of describing the broken-ergodicity phase of the spin glass as a
  multithermal system~\cite{CKM19,CCKM20,K21}. Given these premises,
  what we want to stress here is that for an appropriate choice of
  observables this multithermal description of integrable systems, or
  more in general systems with broken ergodicity, is not necessary.
  Here we want to stress much further this concept by showing that
  thermalization takes place also in integrable systems.
  
   Let us notice that the Hamiltonian of an
  integrable system   takes the form of a sum function, i.e., it can be written as $\mH = \sum_{i=1}^N
  f_i(\mQ_i,\mP_i)$, where $(\mQ_i,\mP_i)$ are the canonical variables
  which allow to diagonalize the Hamiltonian. This is exactly the original case discussed in
  Khinchin's book~\cite{K49}; however it is worth reminding that 
Mazur and van der Linden~\cite{ML63} were able to prove an analogous
result for Hamiltonians which are perturbations of an integrable system.
  }

Bearing in mind the traditional approach of Ma's book~\cite{MA85}, we
adopt the following pragmatic definitions: (a) a system is at
\emph{equilibrium} if the time averages of relevant macroscopic
observables, computed during suitably long time intervals, are in
agreement with the corresponding ensemble averages; (b)
\textit{thermalization} occurs when, starting from an initial
condition far from equilibrium, after a finite time the system reaches
an equilibrium state, as defined above.  To the best of our knowledge,
the choice of the ``relevant'' observables, as already stressed by
Ma~\cite{MA85}, Onsager and Machlup~\cite{OM53}, remains a subtle
point, which cannot be easily bypassed.\\

\mcol{The main message of the paper is that any generic Hamiltonian
  system in the large-$N$ limit exhibits good thermalization and
  equilibrium properties irrespectively to whether or not ergodicity
  holds in the strict mathematical sense, provided that one considers
  ``egalitarian'' observables. By ``egalitarian'' we mean functions
  involving a finite fraction of the degrees of freedom of the system,
  as suggested by Khinchin's approach. In practice, we are going to
  show that the the time averages are well approximated by analytic
  predictions drawn from equilibrium ensembles.}

\section{Model}

We study numerically the Hamiltonian dynamics of the Toda
model~\cite{T67}. The system is constituted by $N$ classical particles
on a line connected by non-linear springs. We design with $q_i$ the
displacement from the equilibrium position of the $i$-th particle,
$i=1, ..., N$, and with $p_i$ its conjugated momentum. The dynamics is
defined by the following Hamiltonian:
\begin{align}
  \mH(q,p) = \sum_{i=1}^N \frac{p_i^2}{2} + \sum_{i=0}^N V(q_{i+1}-q_i),
  \label{eq:HToda}
\end{align}
where $V(x)$ is the Toda potential
\begin{align}
  V(x) = \exp(-x)+x-1,
  \label{eq:Vx}
\end{align}
and we consider fixed boundary conditions, i.e. $q_0=q_{N+1}=0$. It has been shown
by H\'enon in~\cite{He74} that in the Toda lattice there are $N$
conserved quantities, i.e., $N$ functions $\mI_k$ such that $\lbrace \mH,
\mI_k \rbrace = 0$, with $k=1,\ldots,N$; as a consequence, model~\eqref{eq:HToda}
is integrable. The explicit expression of
the first few H\'enon's integral of motion for a periodic system and how
this definition can be adapted to a system with fixed boundary
conditions can be found in Appendix~\ref{app:a1}.

The key idea of this paper is to study indicators of thermal
equilibrium with respect to a set of canonical variables different
from those which diagonalize the Hamiltonian, i.e. the Toda
modes. In particular, we focus our attention on the Fourier modes,
which do not diagonalize the Hamiltonian in Eq.~(\ref{eq:HToda}) but
nevertheless represent a complete basis for any configuration of the
chain. These modes are defined from the set of canonical coordinates
$(q_i,p_i)$ by means of the following sine Fourier transform,
\begin{align}
  \mQ_k &= \sqrt{\frac{2}{N+1}}~\sum_{i=1}^N~q_i~\sin\left( \frac{\pi i k}{N+1}\right) \nonumber \\
  \mP_k &= \sqrt{\frac{2}{N+1}}~\sum_{i=1}^N~p_i~\sin\left( \frac{\pi i k}{N+1}\right),
\end{align}
which represents a canonical change of coordinates. In terms of the
Fourier modes the Hamiltonian in Eq.~(\ref{eq:HToda}) is highly
non-linear. In particular, it has nonlinearities at all orders,
reading as
\begin{align}
  \mH(\mQ,\mP) = \frac{1}{2} \sum_{k=1}^N (\mP^2_k + \omega_k^2
  \mQ^2_k) + \sum_{p=3}^\infty \sum_{k_1,\ldots,k_p} \alpha_{k_1...k_p}~\mQ_{k_1}\cdots\mQ_{k_p}\,,
  \label{eq:Fourier-Hamiltonian}
\end{align}
\textcolor{black}{
where
\begin{equation}
 \omega_k=2\sin \left( \frac{\pi k}{2N+2} \right)
\end{equation}
is the angular frequency of the $k$-th normal mode in the small-energy limit and $\{\alpha_{k_1...k_p}\}$ are suitable coefficients.
}
Let us ignore for a moment that Toda model is integrable: by just looking
at Eq.~(\ref{eq:Fourier-Hamiltonian}), very naively we may conclude
that energy exchanges between Fourier modes are very efficient and
that the system, whatever the initial condition, relaxes to
equilibrium quickly. This is exactly what we find, notwithstanding
integrability. Very roughly, the behavior of the system meets the
most naive expectation that the energy stored in the infinitely many
non-linear terms on the right-hand side of
Eq.~(\ref{eq:Fourier-Hamiltonian}) acts as a thermal bath for the
harmonic terms. We present evidences of this
behavior both studying the relaxation to equilibrium from a very
atypical initial configuration and also studying how phase space is
sampled by the ``equilibrium'' dynamics of the system, i.e., the
Hamiltonian dynamics initialized with a \emph{typical} initial
condition. Clearly, we assume a definition of what is ``typical'' or
``not typical'' having in mind the presumed equilibrium measure,
i.e., the uniform measure on constant energy hypersurfaces.

\section{Relaxational dynamics}

First, we have repeated the celebrated FPUT experiment, i.e. we have
studied how an atypical initial condition, obtained by assigning all
harmonic energy to the lowest mode of the chain, relaxes to
equilibrium, here identified by equipartition among all modes.
\begin{figure}
  \includegraphics[width=0.9\columnwidth]{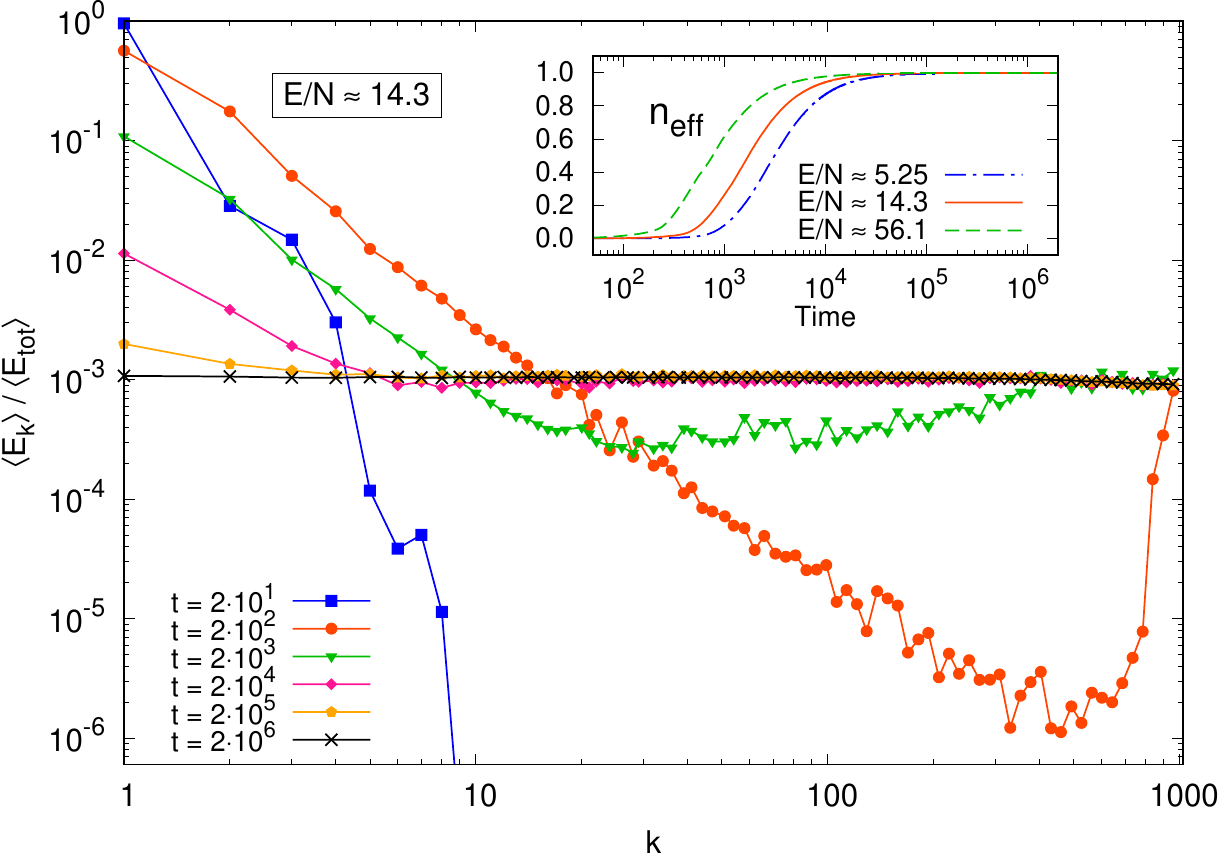}
  \caption{\emph{Main}: Energy spectrum of Fourier modes, $\langle
    E_k\rangle/\langle E_{tot}\rangle$ vs $k$, at different times (different symbols) for a
    Toda lattice with $N=1023$ particles and total energy $E/N \simeq
    14.3$ (here $E_0=4.0$). \emph{Inset}: Effective number of degrees of freedom,
    $n_{eff}$, as function of time for different values of the energy. The value $n_{eff}=1$ signals
    equipartition. Time-step for integration: $\Delta t=0.02$.}
\label{fig1}
\end{figure}
The initial conditions for this non-equilibrium relaxation dynamics
are:
\begin{equation}
  \omega_k^2 Q^2_k = P^2_k =
  \begin{cases}
  E_0~~~~~~k=1\\   
  0~~~~~~~~k\neq 1\,.
  \end{cases}
  \label{eq:init-cond}
\end{equation}
Therefore the value of $E_0$ set in the initial condition does
not corresponds to the total energy of the system, which comprises the
non-linear terms of Eq.~(\ref{eq:Fourier-Hamiltonian}). Along the
non-equilibrium dynamics we have monitored two indicators, one
qualitative the other quantitative. The first is the power spectrum
of the system, 
\begin{equation}
 u_k(t)=\frac{\langle E_k\rangle_t}{\langle E_{tot}\rangle_t}\,,
\end{equation} 
 where $E_k$ denotes the harmonic energy
\begin{align}
  E_k  =  \frac{1}{2}\left[ P^2_k + \omega_k^2 Q^2_k \right]\,,
\end{align}
while $E_{tot}= \sum_{k=1}^N \langle E_k(t)\rangle$ and we consider
cumulative time averages of the kind
\begin{equation}
  \langle E_k \rangle_t  = \frac{1}{t} \int_0^t ds~E_k(s) \,.
  \end{equation}
From Fig.~(\ref{fig1}) it is clear that after a transient where the
harmonic energy remains concentrated on the low-$k$ modes it is then
shared among all modes. Quantitatively this information is encoded in
the behavior of the spectral entropy $S_{\text{sp}}$ and of the
effective number of degrees of freedom $n_{\text{eff}}$, which are
defined as follows:
\begin{align}
  S_{\text{sp}}(t)  &= - \sum_{k=1}^N u_k(t)\log u_k(t) \nonumber \\
  n_{\text{eff}}(t) &= \frac{\exp\left(S_{\text{sp}}\right)}{N}. 
\end{align}
The behavior of $ n_{\text{eff}}(t)$ is shown in the inset of
Fig.~\ref{fig1}: we find the typical sigmoidal shape starting from
small values at the initial times, where energy is localized in the
spectrum, and approaching $n_{\text{eff}}=1$, the value typical of
equipartition at later times. We thus see that, notwithstanding the fact that all
Lyapunov exponents are zero, the system exhibits an \emph{emerging}
relaxational timescale. In the main panel of Fig.~\ref{fig1} we
show results for a given value of the total energy $E$; from the behavior of
$n_{\text{eff}}(t)$ for various choices of the total energy $E$ in the inset
it is clear that the qualitative feature does not change
at varying $E$. Problems such as the dependence of the characteristic
timescale on the initial energy and more in general the dependence on
the initial condition and on the system size $N$ are of course
technically interesting, but are left for future
investigations.\\

\begin{figure}
  \includegraphics[width=0.9\columnwidth]{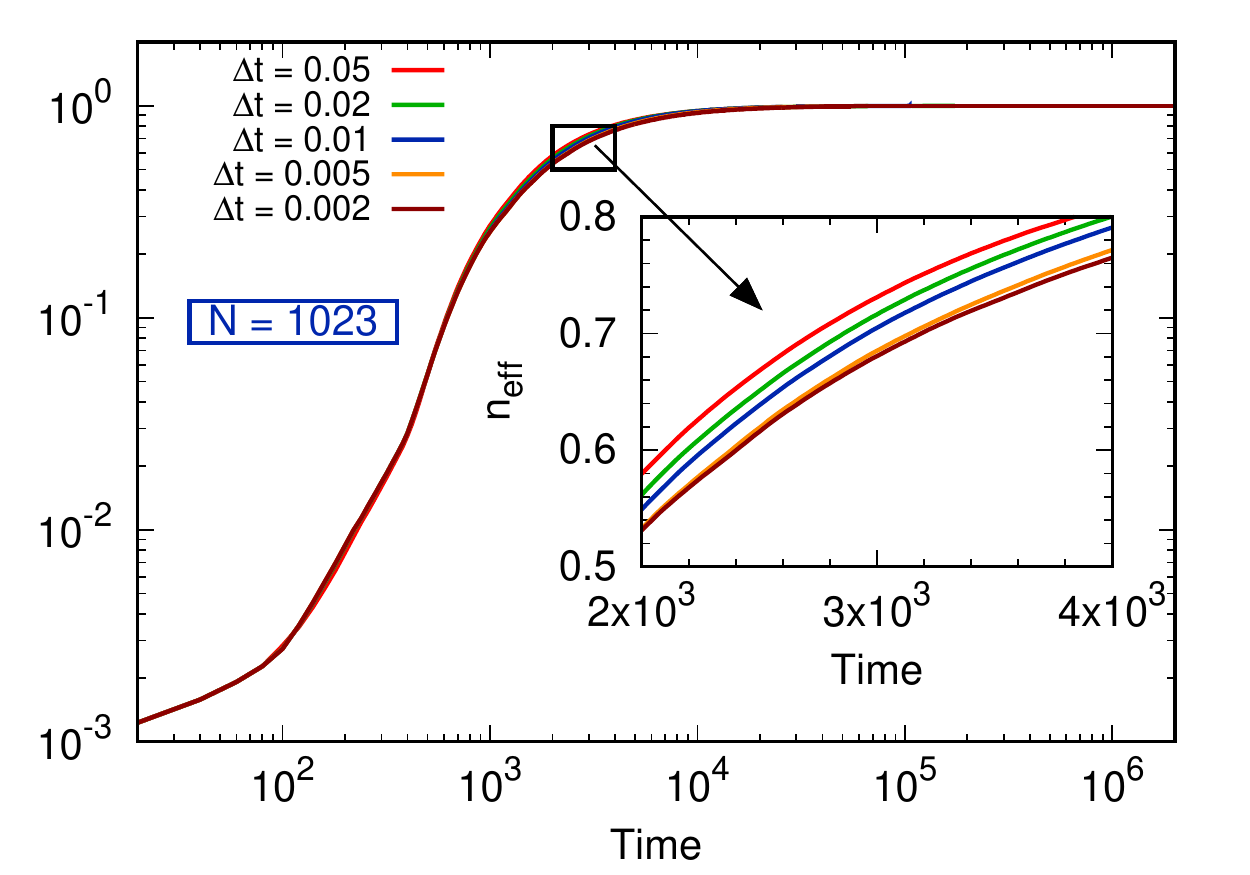}
  \caption{\ggcol{\emph{Main}: Effective number of degrees of freedom
      as a function of time, $n_{eff}(t)$ vs $t$, for different
      choices of timestep $\Delta t$ in the numerical integration
      algorithm. All results are for the system size $N=1023$.
      \emph{Inset}: Zoom of the curves.}}
\label{fig1b}
\end{figure}

\ggcol{Before presenting other tests of thermalization let us comment
  on the reliability of numerics, usually characterized by unavoidable
  approximations such as the study of finite-size systems, finite
  lapses of time and discrete time dynamics. In fact, there are two
  sources of troubles which cannot be avoided in the numerical
  integration of differential equations: (a), the algorithm for the
  integration from time $t$ to $t+\Delta t$ cannot be exact; (b), the
  computer necessarily works with integer numbers. Numerical simulations
  of dynamical systems cannot achieve arbitrary precision. Therefore, even if the system is
  strictly deterministic, as is the case of Hamiltonian systems, its
  numerical study is always affected by a small amount of
  ``randomness'' due to numerical roundings. Nevertheless, it has been
  shown that symplectic algorithms  are rather efficient in limiting
  the spreading of numerical errors in the simulation of Hamiltonian
 systems; this is related to the (important) fact that, at
variance with other popular algorithms (e.g. Runge-Kutta) they allow to
preserve the Hamiltonian structure during the time evolution. For this reason in our numerical
  simulations we used the well-known leap-frog symplectic
  integrator~\cite{Y90}.\\

  To test the stability of numerical results is a compelling issue, in
  particular for the conclusions drawn in this paper. Fast
  equipartition between Fourier modes starting from very atypical
  initial configurations, although in agreement with the premises of
  Khinchin's approach, is a rather unexpected result for an integrable
  system. Usually non-linear Hamiltonian systems are characterized by
  long time-scales emerging in the proximity of an integrable
  limit~\cite{G05,BCP13}. And in general, for classical integrable
  systems, one expects signatures of integrability to be manifest in
  every cirmustance~\cite{S20,D90} (not even to mention the case of
  integrable quantum systems~\cite{VR16,dLM16,SBDD19}). A study of how
  equipartition times depend on the numerical protocol is therefore
  mandatory in the present situation. In order to check that
  equipartition between Fourier modes is not an artifact of the
  numerical protocol we have verified that our results do not depend
  on the time-step of the algorithm. This has been done for a chain of
  $N=1023$ particles. In particular in Fig.~\ref{fig1b} we show that
  varying $\Delta t$ does not lead to any sensible increase of
  relaxation time. This result is not a proof, of course, but it gives
  us a very reasonable argument to expect that thermalization also
  occurs in the infinite precision limit, $\Delta t \rightarrow 0$.\\

  Apart the above practical test, in order to support the validity of
  our conclusions on thermalization it is worth to recall a result
  demonstrated mathematically only for Anosov system but accepted in
  more generic situations. What has been shown in~\cite{BCGGS78,CFV89}
  is that for a high precision integration algorithm, even in the case
  of arbitrary deviations of the numerical trajectories from the
  ``true'' ones (same initial conditions) due to rounding errors, the
  dynamical averages performed along the ``original'' trajectory and
  the one perturbed by rouding errors do coincide.  This means that,
  assuming some weak form of ergodicity, the ``statistical
  reproducibility'' of the true dynamics by the numerical simulations
  is typical~\cite{BCGGS78,CFV89}.}

\section{Equilibrium dynamics}

\begin{figure}
  \includegraphics[width=0.9\columnwidth]{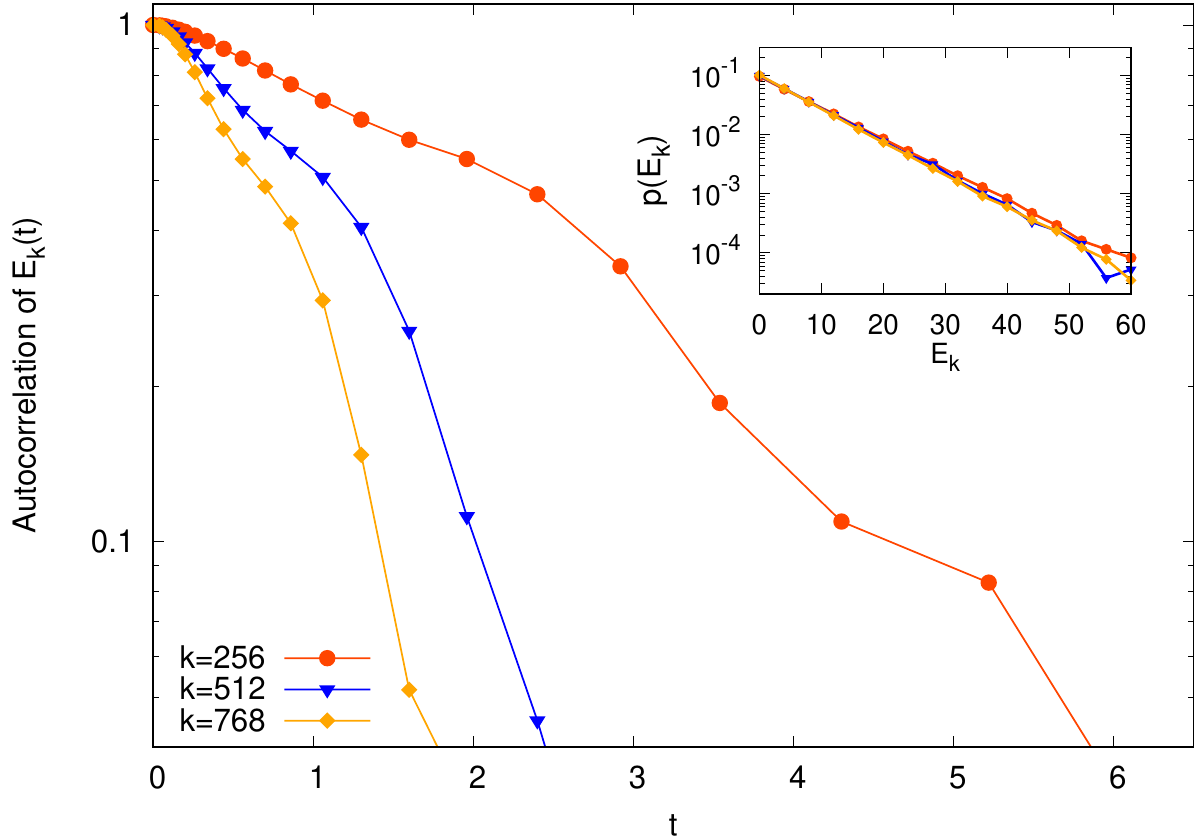}
  \caption{\emph{Main}: Time correlation function for the harmonic
    energy at different values of $k$: $k=256,512,768$ for a Toda
    lattice with $N=1023$ particles and energy $E/N =
    4.0$. \emph{Inset}: Probability distribution of the harmonic
    energy $\varepsilon_k$ for different values of $k$.}
\label{fig2}
\end{figure}

Initial conditions which are both good to mimic thermal equilibrium
and to set precisely the initial value of the energy are obtained by
putting all initial amplitudes $Q_k$ equal to zero and taking
\begin{align}
    \sum_{k=0}P^2_k = 2E_0
  \label{eq:init-cond-eq}  
\end{align}
This condition can be obtained quite easily by sampling the initial
velocity of each particle from a Gaussian distribution, which
guarantees a good degree of randomization, and then rescaling all
velocities according to the chosen value of $E_0$. One then considers
equilibrium observables such as the time decay of the harmonic
energies self-correlations:
\begin{align}
 C_k(t) = \frac{\langle E_k(t) E_k(0)\rangle - \langle E_k\rangle^2}{\langle E_k^2\rangle-\langle E_k\rangle^2}\,,  
 \label{eq:ckt}
\end{align}
and the probability distribution of the harmonic energy of a given
mode $p(E_k)$, which is simply defined as the histogram of values
taken by $E_k$. In Fig.~\ref{fig2} it is shown that $C_k(t)$, for some
given values of $k$, rapidly decays to zero. At the same time we find
a nice exponential behavior for $p(E_k)$:
\begin{align}
  p(E_k) \sim \exp\left( -b E_k\right),
  \label{eq:pek}
\end{align}
where the value of $b$ is close to the inverse of the temperature in
the simulation at small energies while it deviates at high
energies. The description of the system in terms of normal modes is
indeed appropriate for energy not too large while in the opposite
limit of very high energy the system becomes similar to hard
spheres.\\
\begin{figure}
  \includegraphics[width=0.9\columnwidth]{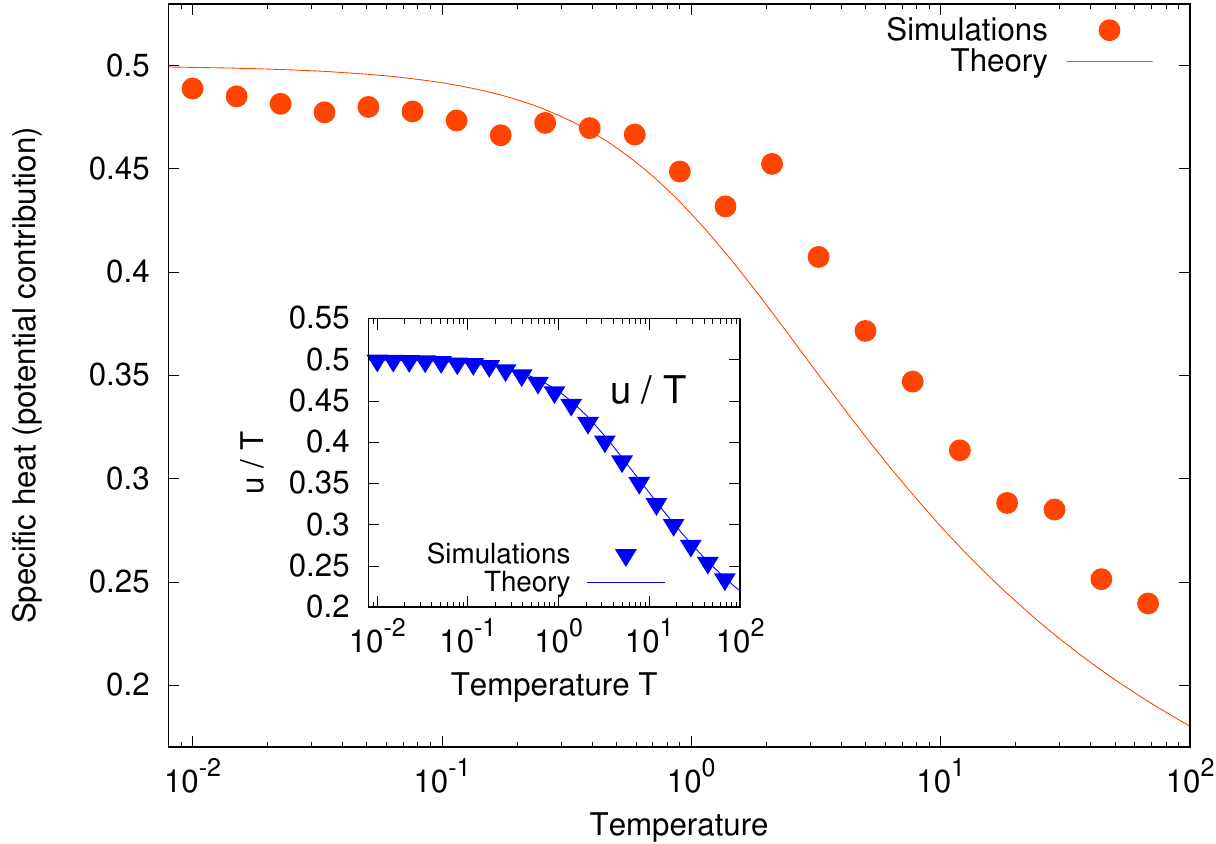}
  \caption{\emph{Main}: Potential contribution of the specific heat $C_V$ as a function of
    temperature $T$. Points: numerical data obtained by using
    Eq.~\eqref{eq:spheat} with $N_*=50$ and averaging over 10
    different subsystems, for a total time $\mathcal{T}=5\cdot10^6$; lines: analytical
    prediction.  \emph{Inset}: average \emph{potential} energy as a
    function of temperature $T$; points, numerical data; lines, analytical
    prediction.}
\label{fig3}
\end{figure}

Let us now compare the time average of some observables of obvious
interest computed in simulations with initial conditions as in
Eq.~(\ref{eq:init-cond-eq}) with the prediction which can be worked
out explicitly from the calculation of the partition function
\begin{align}
  \mathcal{Z}_N(\beta) = \mathcal{Z}_N^{(\text{K})}(\beta) \cdot \mathcal{Z}_N^{(\text{P})}(\beta),
\end{align}
where the kinetic and the potential contributions, respectively
$\mathcal{Z}_N^{(\text{K})}(\beta)$ and
$\mathcal{Z}_N^{(\text{P})}(\beta)$, are factorized and read
respectively as
\begin{align}
  \mathcal{Z}_N^{(\text{K})}(\beta) &= \beta^{ - N/2} \nonumber \\
  \mathcal{Z}_N^{(\text{P})}(\beta) &= \int_{-\infty}^\infty \prod_{i=1}^{N+1} dr_i~e^{-\beta V(r_i)}~\delta\left( \sum_{i=1}^{N+1} r_i \right) \nonumber \\
  & = e^{\beta N}\int_{s_0-i\infty}^{s_0+i\infty} \frac{ds}{2\pi i}~\exp\left\lbrace N \log z_\beta(s) \right\rbrace,
\end{align}
with
\begin{align}
z_\beta(s) = \int_{-\infty}^\infty dr~\exp\left(-\beta e^{-r} + s r\right).
\end{align}
The partition function $\mathcal{Z}_N^{(\text{P})}$ can be computed
exactly in the large-$N$ limit (details can be found in~App.~\ref{app:a2}),
yielding:
\begin{align}
  \mathcal{Z}_N^{(\text{P})} = e^{\beta N}~\beta^{-\mu(\beta)}~\Gamma[\mu(\beta)],
  \label{eq:exact-Z}
\end{align}
where the function $\mu(\beta)$ is determined implicitly by the
condition:
\begin{equation}
  \frac{\Gamma'(\mu)}{\Gamma(\mu)} = \log(\beta),
  \label{eq:sp-condition}
\end{equation}
and does not have in general an explicit expression.  For the
comparison between simulation and theory let us now focus \emph{only}
on the potential energy partition function
$\mathcal{Z}_N^{(\text{P})}(\beta)$. By plugging the expression of
$\mathcal{Z}_N^{(\text{P})}(\beta)$ from Eq.~(\ref{eq:exact-Z}) into the
standard formulas
\begin{align}
\langle u \rangle & = - { 1 \over N}  \frac{\partial}{\partial\beta}\log \mathcal{Z}_N^{(\text{P})}(\beta) \nonumber \\
C_V & =  { \beta^2 \over N}  \frac{\partial^2}{\partial\beta^2}\log \mathcal{Z}_N^{(\text{P})}(\beta),
\label{eq:em-CV-formula}
\end{align}
one obtains the analytical predictions for the average potential energy $\langle
u \rangle$ and the specific heat $C_V$ which are plotted in
Fig.~\ref{fig3}, where they are also compared with the numerical
data. Apart from small discrepancies for $C_V$, probably due to the
lack of large enough statistics, it is clear that the numerical
behavior is well reproduced by the exact computation of $C_V$ done
assuming thermal equilibrium in the canonical ensemble. The formula of
$C_V$ cannot be given in the general case explicitly in terms of
$\beta$, because one needs the knowledge of $\mu(\beta)$, which is
determined only implicitly by Eq.~(\ref{eq:sp-condition}). An
expression of $C_V$ in terms of $\mu(\beta)$ can be found
in~App.~\ref{app:a2}. What can be written explicitly are the asymptotic
behaviors, which read:
\begin{align}
  T \rightarrow 0 ~~~&\Longrightarrow~~~\langle u \rangle = \frac{T}{2}; ~~~~~~~~~C_V = \frac{1}{2} \nonumber \\
  T \rightarrow \infty ~~~&\Longrightarrow~~~\langle u \rangle\sim \frac{T}{\log(T)};~~~ C_V \sim \frac{1}{\log(T)}\,.
\end{align}
Asymptotically, we have
\begin{align}
C_V  \approx \langle u \rangle/T,
\end{align}
which nicely compares with the results shown in Fig.~\ref{fig3}. We
have thus shown that, despite the Toda model being integrable, its
behavior is strikingly well captured by equilibrium statistical
mechanics formulas. In order to simulate a canonical ensemble avoiding
any noise source modeling the heat bath, we have just studied the
Hamiltonian dynamics of a systems with $N$ particles and focused our
attention on subsystems of $N_*$ particles each, with $1 \ll N_* \ll
N$. For instance we have considered the subsystems in the central part
of the chain, in such a way that the fixed boundary conditions are not very
relevant. In the simulations the specific heat has been computed as
\begin{align}
\label{eq:spheat}
C_V={\langle E_*^2\rangle -\langle E_* \rangle^2 \over N_* T^2},
\end{align}
where $E_*$ is the energy in the subsystem with $N_*$ particles and
temperature is defined as $T = \langle p^2 \rangle$.\\

Let us stress that the good agreement between equilibrium
calculations and numerical simulations is not particularly
surprising for what concerns the behavior of the average energy
$\langle u \rangle$: basically, it is a sort of virialization of the
global kinetic energy starting from initial conditions which are not
at equilibrium.  On the contrary the test for the specific heat, which
involves the behavior of the energy fluctuations of a subsystem, is
rather severe. Let us note that this result for the specific heat is
quite analogous to the one obtained in the FPUT system~\cite{LPRV87},
where chaos is present.\\

\section{Conclusions}

We have shown, by comparing analytical results with numerical
simulations, that even the dynamics of an integrable system such as
the Toda chain has good ``equilibrium'' properties, provided one looks
at ``generic'' observables. In the present context a good definition
of ``generic'' looks as the one of \emph{``collective with respect to
  the canonical coordinates which diagonalize the Hamiltonian''}. This
observation suggests the possibility to generalize the implications of
our results to \ggcol{generic quantum systems, i.e. to the
  \emph{non-integrable} quantum unitary dynamics.  The Schr\"odinger
  equation has in fact, by construction, properties very similar to
  integrable systems. In particular, the coherent motion of the
  eigenvectors phases is analogous, in the case of a $N$-dimensional
  system, to the one of the $N$ angles of the Liouville-Arnold's
  theorem action-angle pairs. The analogy between \emph{generic}
  quantum systems and integrable classical ones stems from the
  conservation along quantum dynamics of the $N$ projectors on energy
  eigenstates, $\hat{P}_\alpha = |\varepsilon_\alpha \rangle \langle
  \varepsilon_\alpha |$. We do not consider here the more subtle case
  of \emph{integrable} quantum systems, where these $N$ global
  conservation laws are also accompanied by the conservation of $N$
  local charges $Q_i$. Finding what is the ``classical analog'' of an
  integrable quantum system is therefore a subtler issue: one
  needs to find a system with $2N$ conservation laws, among which $N$
  are global and $N$ are local. On the other hand, the problem of
  thermalization is interesting already in the case of non-integrable
  quantum systems. For instance in the case of small systems sizes,
  where the validity of ``bona-fide'' ergodicity assumptions as the
  Eigenstate Thermalization Hypothesis is not granted~\cite{RS12,D18},
  one finds clear signatures of the built-in integrability due to
  quantum unitary dynamics~\cite{O15}.

  If we thus try to extrapolate the implications of our results to
  quantum systems, we may say that the decoherence between the phases
  of different eigenvectors may be \emph{not necessary} for the
  thermalization of the system: to detect thermalization one just
  needs to look at observables uniformly spread on \emph{all}
  eigenvectors. This last conclusion seems quite in line with Von
  Neumann's ``quantum ergodic theorem''~\cite{JVN29,GLTZ10}.
  According to the latter, relaxation to a thermal state is a property
  of the chosen observables rather than of the energy eigenmodes
  (absence of) structure. Structureless eigemodes are in fact the
  landmark of the so-called ``quantum chaos''~\cite{BGS84}. A similar
  thing happens also for the Eigenstate Thermalization Hypothesis,
  which is regarded by someone as the modern version of the Von
  Neumann theorem. Then, what is similar between the Von Neumann
  quantum ergodic theorem and the results due to Khinchin, Mazur and
  van der Linden is the lack of any explicit mention to quantum or
  dynamical chaos. Our numerical results on the Toda model seem to be
  quite in line with the idea of \emph{``thermalization without
    chaos''}. Further investigations are for sure needed to clarify
  the problem of thermalization in the presence of conservation laws
  in high-dimensional Hamiltonian systems, the study of thermal
  properties in isolated quantum systems and classical integrable ones
  is a topic which keeps on attracting many scholars, see
  also~\cite{GLTZ10,CK12,BCLN20}}.\\ \\



{\bf Acknowledgments}~\ggcol{We acknowledge for useful exchanges A. De Luca, T. Goldfriend,
J. Kurchan, R. Livi, G. Mussardo, A. Ponno, V. Ros, A. Scardicchio.} M.B. and
A.V. acknowledge partial financial support of project MIUR-PRIN2017
\textit{Coarse-grained description for non-equilibrium systems and
  transport phenomena (CO-NEST)}.


\newpage

\setcounter{section}{0}

\section{Appendix}

\subsection{Integrals of Motion}
\label{app:a1}

Just for the sake of self-consistency we briefly remind here the
conservation laws of the Toda system. In the case of a periodic
lattice with $N$ particles where $q_0=q_N$ the first four integrals of
motion read as:
\begin{align}
  \mI_1 &= \sum_{i=1}^N p_i \nonumber \\
  \mI_2 &= \sum_{i=1}^N ~\frac{p_i^2}{2} + \mU_i \nonumber  \\
  \mI_3 &= \sum_{i=1}^N ~\frac{p_i^3}{3} + (p_i+p_{i+1})\mU_i \nonumber \\
  \mI_4 &= \sum_{i=1}^N ~\frac{p_i^4}{4} + (p_i^2+p_{i+1}^2+p_ip_{i+1})\mU_i + \frac{\mU_i^2}{2}+\mU_i\mU_{i+1}\nonumber, \\
  \label{eq:toda-integrals}
\end{align}
where $\mU_i = e^{-(q_{i+1}-q_i)}$. Note that $\mI_2$ is nothing but
the energy (apart for an irrelevant additive constant). The expressions in Eq.~(\ref{eq:toda-integrals}), see,
e.g.,~\cite{He74}, can be extended to the case of fixed boundary
conditions by conventionally defining a periodic chain with $2N+2$
particles where the coordinates from $q_0$ to $q_{N+1}$ are identical
to the original one, while those such as $q_{N+1+i}$ with
$i=1,\ldots,N$ are defined in order to have an odd function with
respect to the center of the lattice:
\begin{align}
  q_{N+1+i} &= -q_{N+1-i} \nonumber \\
  p_{N+1+i} &= -p_{N+1-i} \nonumber \\
\end{align}

\subsection{Specific Heats}
\label{app:a2}

The first step to compute the specific heat is the calculation of the
partition function. For consistency with the numerical simulations,
we compute it with fixed boundary conditions. That is, we have
$N$ variables $q_i$ with $i=1,\ldots,N$ to integrate over, while
\be 
q_0 = q_{N+1} = 0 
\ee
By identifying with the same variable the two fixed boundaries,
i.e., $q_0=q_{N+1}$, we can thus change variables to
\be
 r_i = q_{i+1} - q_i,
\ee
with, in particular
\begin{align}
  r_{N} &= q_{N+1} - q_N, \nonumber \\
  r_{N+1} &= q_1 - q_{N+1}. 
\end{align}
We can thus safely integrate over the $N+1$ variables $r_i$ under the
global constraint $\sum_{i=1}^{N+1}r_i = 0$:
\begin{align}
\mathcal{Z}_N^{(\text{P})}(\beta) = e^{\beta N} \int_{-\infty}^\infty
\prod_{i=1}^{N+1} dr_i~e^{-\beta \sum_{i=1}^{N+1} \exp(-r_i)}~\delta\left(
\sum_{i=1}^{N+1} r_i \right).
\end{align}
In order to \emph{unfold} the global constraint it is useful to
exploit the inverse Laplace rapresentation of the partition function:
\begin{align}
  & \mathcal{Z}_N^{(\text{P})}(\beta) = \nonumber \\
  & = e^{\beta N}\int_{s_0-i\infty}^{s_0+i\infty} \frac{ds}{2\pi i} \int_{-\infty}^\infty \prod_{i=1}^{N+1} dr_i~
  \exp\left( -\beta \sum_{i=1}^{N+1} \exp(-r_i) + s \sum_{i=1}^{N+1} r_i \right) \nonumber \\
  & = e^{\beta N}\int_{s_0-i\infty}^{s_0+i\infty} \frac{ds}{2\pi i}
  \left[\int_{-\infty}^\infty dr~\exp\left(-\beta e^{-r} + s r\right) \right]^{N+1} \nonumber \\
  & =  e^{\beta N}\int_{s_0-i\infty}^{s_0+i\infty} \frac{ds}{2\pi i}~\exp\left\lbrace N \log z_\beta(s) \right\rbrace
\label{eq:Z-Laplace}
\end{align}
where we have introduced the partition function per degree of freedom:
\be
z_\beta(s) = \int_{-\infty}^\infty dr~\exp\left(-\beta e^{-r} + s r\right).
\label{eq:z}
\ee
It is convenient to regard $\beta$ as a parameter in Eq.~(\ref{eq:z})
and consider $z_\beta(s)$ as a function of $s$ in the complex $s$
plane. Clearly the integral expression in Eq.~(\ref{eq:z}) is well
defined only for $\text{Re}(s)<0$, so that it is convenient to change
variable in the integral of Eq.~(\ref{eq:Z-Laplace}) from $s$ to $\mu = -s$:
\be
\mZ_N^{(\text{P})}(\beta) = e^{\beta N}\int_{\mu_0-i\infty}^{\mu_0+i\infty} \frac{d\mu}{2\pi i}~
\exp\left\lbrace N \log \hat{z}_\beta(\mu) \right\rbrace,
\label{eq:Z-new-L}
\ee
where now the contour in the complex plane passes through
$\text{Re}(\mu_0) > 0$ and we have the partition function
per degree of freedom defined as:
\be
\hat{z}_\beta(\mu) = \int_{-\infty}^\infty dr~\exp\left(-\beta e^{-r} - \mu r\right).
\label{eq:zeta-mu}
\ee
The partition function of the whole system $Z_N(\beta)$ can be
computed from Eq.~(\ref{eq:Z-new-L}) by means of a saddle-point
approximation in the large-$N$ limit. But first we need to compute
explicitly $\hat{z}_\beta(\mu)$ in Eq.~(\ref{eq:zeta-mu}). By going through
two changes of variables, i.e.,
\begin{align}
  x\rightarrow u &= e^{-x} \nonumber \\
  u \rightarrow t &= \beta u,
\end{align}
we obtain
\begin{align}
  \hat{z}_\beta(\mu) &= \int_0^\infty du~u^{\mu-1}e^{-\beta u} = \frac{1}{\beta^\mu} \int_0^\infty dt~t^{\mu-1}~e^{-t} \nonumber \\
  \hat{z}_\beta(\mu) & = \frac{\Gamma(\mu)}{\beta^\mu}.
\end{align}
In order to compute the integral in Eq.~(\ref{eq:Z-new-L}) by means of
the saddle-point approximation we just need to solve the saddle-point
equation
\begin{align}
  & \frac{\partial}{\partial \mu}\left(\log \Gamma(\mu) - \mu \log(\beta) \right) = 0 \nonumber \\
  & \Longrightarrow~\frac{\Gamma'(\mu)}{\Gamma(\mu)} = \log(\beta)
  \label{eq:saddle-point}
\end{align}
At this stage, without doing any approximation and indicating as
$\hat{\mu}(\beta)$ the function implicitly determined by the condition
in Eq.~(\ref{eq:saddle-point}), one has that the partition function of
the system reads:
\begin{align}
  \mZ_N^{(\text{P})}(\beta) = e^{\beta N} \frac{\Gamma(\hat{\mu}(\beta))}{\beta^{\hat{\mu}(\beta)}}.
  \label{eq:Zeta-mu}
\end{align}
For instance the specific heat $C_V$, which can be obtained from
Eq.~(\ref{eq:Zeta-mu}) using the standard formula in
Eq.~(18) of the main text, reads as
\begin{widetext}
\begin{align}
  C_V = - \frac{1}{\psi'(\hat{\mu}(\beta))} - \left[ \psi(\hat{\mu}(\beta))- \log(\beta) \right]
  \left( \frac{\psi^{''}(\hat{\mu}(\beta))}{[\psi'(\hat{\mu}(\beta))]^3} + \frac{1}{\psi'(\hat{\mu}(\beta))} \right) + \hat{\mu}(\beta),
  \label{eq:Cv-complete}
\end{align}
\end{widetext}
being $\psi(x)$ the so-called \emph{digamma} function $\psi(x) =
\Gamma'(x)/\Gamma(x)$. The expression in Eq.~(\ref{eq:Cv-complete}) is
not very insightful, but is necessary to plot the full analytic
behavior of the function. Similarly one can write down the expression
for the temperature, but we leave this exercise to the willing
reader.\\

On the contrary, what can be obtained explicitly, are the asymptotic
behaviors for both the average potential energy $\langle u \rangle$
and the specific heat $C_V$ in the limit of small and large
temperatures $T$. This is obtained by the knowledge of the two
asymptotic behaviors of the digamma function $\psi(\mu)$:
\begin{align}
  \mu \rightarrow \infty ~~~&\Longrightarrow~~~\psi(\mu) \sim \log(\mu) \nonumber \\
  \mu \rightarrow 0~~~&\Longrightarrow~~~\psi(\mu) \sim - \frac{1}{\mu}. 
\end{align}
We thus have two different approximations for our saddle-point
equations in the \emph{low-temperature}, $\beta \rightarrow \infty$,
or the \emph{high-temperature}, $\beta \rightarrow 0$, limit:
\begin{align}
  \text{Low T:}~~~\Longrightarrow~~~ \log(\mu) &= \log(\beta)~~~\Longrightarrow~~~\mu=\beta \nonumber \\
  \text{High T:}~~~\Longrightarrow~~~ -\frac{1}{\mu} &= \log(\beta)~~~\Longrightarrow~~~\mu=-\frac{1}{\log(\beta)} \nonumber \\
\end{align}
Let us first consider the low-temperature case, where Toda is
well-approximated by a system of weakly coupled harmonic oscillators,
for which the specific heat is constant. In this case we get:
\begin{align}
Z_N(\beta) \approx \exp\left( (N+1) \left[ \beta + \log\Gamma(\beta) -\beta\log(\beta) \right] \right).
\end{align}
By using the Stirling's approximation
\be
\log\Gamma(\beta) = \beta\log(\beta) - \beta -\frac{1}{2} \log(\beta), 
\ee
we thus get
\be
Z_N(\beta) \approx \exp\left[ - \frac{N}{2}\log(\beta) \right].
\label{eq:Z-largeb}
\ee
From Eq.~(\ref{eq:Z-largeb}) and from the definition of $C_V$ in
Eq.~(18) in the main text we get
\be
C_V = \beta^2\frac{\partial^2}{\partial \beta^2}\left( \frac{N}{2}\log(\beta) \right) = \frac{1}{2}.
\ee
In the light of the high-temperature expansion, we get
\be
\beta \rightarrow 0 \Longrightarrow 
\mathcal{Z}_N(\beta)
\approx\left\lbrace N\left[ \beta +\log \Gamma\left( -\frac{1}{\log(\beta)}\right)\right] \right\rbrace \nonumber. \\
\ee
By exploiting the identity of the Euler gamma function
$z\Gamma(z)=\Gamma(z+1)$ one has that
\begin{align}
z \rightarrow 0 ~~~\Longrightarrow~~~\Gamma(z) \sim \frac{1}{z},
\end{align}
from which we have
\begin{align}
 &\log\mathcal{Z}_N(\beta)
  \approx\left\lbrace N\left[ \beta +\log\left(-\log(\beta)\right)\right] \right\rbrace \nonumber \\
 & \Longrightarrow ~~~C_V \sim  \frac{1}{\log(T)}
\end{align}
In a similar manner one obtains the behavior of the average energy
$\langle u \rangle$, which is reported in the main text.

\end{document}